\documentstyle[a4,11pt]{article}
\def\slash{\!\!\!/}
\begin{document}     
\title{{\bf A Comment on Heavy Quark Effective Fields}
\vspace*{0.2in}
\thanks{Research partially supported by CICYT under contract AEN-96/1718}}
\author{Miguel Angel Sanchis-Lozano \thanks{E-mail: mas@evalvx.ific.uv.es} \\
\\
\it Departamento de F\'{\i}sica Te\'orica and IFIC\\
\it Centro Mixto Universidad de Valencia-CSIC\\
\it 46100 Burjassot, Valencia, Spain}
\maketitle 
\vspace*{1.0in}
\abstract{Effective fields defined in the heavy-quark effective theory
to describe heavy quarks in heavy-light hadrons are examined
in some detail in the standard formulation of a quantum field theory.}
\large
\vspace{-13.cm}
\begin{flushright}
  FTUV/96-47 \\
  IFIC/96-55
\end{flushright} 
\vspace{10.cm}
\newpage
Over the past decade, the approach to the weak decays of heavy (D and B)
mesons can be characterized by the birth of more or less sophisticated
phenomenological models 
\cite{alta} dealing with the complexity of the strong interaction dynamics. 
On the other hand, the current decade has witnessed the development
of effective theories for the strong interaction (like the heavy-quark 
effective theory or HQET \cite{grins} \cite{neub}) coming from
first principles, allowing very important phenomenological applications 
\cite{neu}. However, attention should still be
paid to some formal aspects of the HQET as a proper quantum 
 field theory as literature is sometimes
confusing in this \vspace{0.1in} regard.
\par
In this Letter, I shall examine some definitions for
the effective fields describing heavy quarks according to
the HQET, working out Feynman propagators from them. In particular
I shall analyze their Fourier content in momentum space
in the context of a quantum field \vspace{0.1in} theory.
  
\par 
Let us suppose a generic hadron moving with four-velocity $v$
$(v^2=1)$,  
made of a heavy quark $Q$ along with a light component. According to
Georgi's original remarks \cite{geor} the heavy quark becomes
cannonball conserving its velocity, almost the same as the
hadron's, unaltered as
long as it interacts softly with the light degrees of freedom - without
undergoing any weak or strong decay. Thereby, the heavy
quark velocity can be viewed as a new {\em label} for heavy quarks, like
 \vspace{0.1in} flavour.
\par
It is customary to begin any pedagogical
introduction to HQET by expressing the heavy quark four-momentum $p_Q$
as the sum $p_Q=m_Qv+k_Q$ where $m_Qv$ represents the
(large) mechanical momentum whereas $k_Q$ stands for
the residual momentum of the almost on-shell heavy quark inside the
\vspace{0.1in} hadron \footnote{Subindex $Q$ has been added to those
quantities like $p_Q$ or $k_Q$ to be interpreted as operator
expectation values for a particular hadron state, satisfying 
the relation $v\ {\simeq}\ p_Q/m_Q$. All the components of $k_Q$ are
of the order of ${\Lambda}_{QCD}$ and we shall express the
off-shellness of the heavy quark as
$p_Q^2=m_Q^2+2m_Qv{\cdot}k_Q+k_Q^2=m_Q^2+{\Delta}^2$. In the text we 
substitute the index $Q$ by $b$, which may be identified for most
realistic applications as the bottom flavour in particular}. 
\par
Let us start by writing the plane wave Fourier expansion for the 
fermionic field of an almost on-shell heavy quark $Q_v^{(b,\alpha)}(x)$
of flavour $b$ and colour $\alpha$, moving inside a hadron
with four-velocity \vspace{0.1in} $v$ 
\begin{eqnarray}
Q_v^{(b,\alpha)}(x) & = & Q_v^{(b,\alpha)+}(x)\ +\ 
Q_v^{(b,\alpha)-}(x) \nonumber \\
           & = & \int \frac{d^3\vec{p}}{J}\ \sum_{r}\ 
[\ b_r^{\alpha}(\vec{p})\ u_r(\vec{p})\ e^{-ip{\cdot}x}+ 
\ \tilde{b}_r^{\alpha\dag}(\vec{p})\ 
v_r(\vec{p})\ e^{ip{\cdot}x}\ ]
\end{eqnarray}
where $r$ refers to the spin and $J$ stands for the chosen
normalization; $b_r^{\alpha}(\vec{p})/\tilde{b}_r^{\alpha\dag}(\vec{p})$
is the annihilation/creation operator for a heavy quark/antiquark
with three-momentum $\vec{p}$ ($p^0{\simeq}+\sqrt{m_b^2+\vec{p}^2}$). 
Let us firstly consider the particle sector of the \vspace{0.2in} theory.
\newline
{\bf Particle \vspace{0.1in} Sector}
\par
As the heavy quark is almost on-shell we shall require for
{\em each Fourier component} that
\begin{equation}
p^2\ =\ m_b^2\ +\ {\Delta}^2
\end{equation}
where ${\Delta}^2$ is independent of $\vec{p}$ and satisfies
\vspace{0.1in} 
${\lim}_{m_b{\rightarrow}\infty}{\Delta}^2/m_b^2\ {\rightarrow}\ 0$.
\par
On the other hand, spinors are normalized such that
$u_r^{\dag}(\vec{p})u_s(\vec{p})=2p^0N\ {\delta}_{rs}$ and the 
creation/annihilation operators satisfy: 
\[ [b_r^{\alpha}(\vec{p'}),b_s^{\beta\dag}(\vec{p})]_+=
K\ {\delta}^{\alpha\beta}\ {\delta}_{rs}\ {\delta}^3(\vec{p'}-\vec{p}) \]
where $K,\ N$ are the corresponding normalization \vspace{0.1in} factors.
\par
Let us redefine the momentum of each Fourier component in Eq. (1)
according to HQET as the sum of a mechanical part and
a {\em Fourier residual four-momentum} $k$, 
\begin{equation}
p\ =\ m_bv\ +\ k
\end{equation}
\par
Hence one may write  
\begin{equation}
Q_v^{(b,\alpha)+}(x)\ =\ e^{-im_bv{\cdot}x}\ \int\  
\frac{d^3\vec{k}}{J}\ \sum_r\ b_r^{\alpha}(\vec{k})\ u_r(\vec{k})\ 
e^{-ik{\cdot}x}
\end{equation}
\par
The main point to be stressed is that we shall require that each
Fourier component should satisfy the almost on-shell condition (2). 
Therefore, in the hadron rest frame $k{\cdot}v=k^0$ is related to
$\vec{k}$ through the constraint 
\begin{equation}
(k^0)^2+2m_bk^0-\vec{k}^2\ =\ {\Delta}^2
\end{equation}
which yields the expected relation
\begin{equation}
k^0\ =\ {\pm}\sqrt{m_b^2+\vec{k}^2+{\Delta}^2}\ - m_b
\end{equation}
where the positive solution can be identified as approximately
the kinetic energy of the heavy quark inside the hadron; 
the negative solution can be rejected as yields a
negative $p^0$. It is also interesting to note the space-like character of 
\vspace{0.1in} $k$.
\par
Notice that the annihilation (and creation) operators and spinors 
have been simply {\em relabelled} in Eq. (4), satisfying the same 
normalization as above, though expressed in terms of the Fourier 
residual momentum $\vec{k}$. 
In particular, $b_r^{\alpha}(\vec{k})/b_r^{\alpha\dag}(\vec{k})$
corresponds to annihilation/creation operators
for a heavy quark with residual momentum $\vec{k}$ in a hadron
moving with four-velocity \vspace{0.1in} $v$, satisfying accordingly

\begin{equation}
[b_r^{\alpha}(\vec{k}'),b_s^{\beta\dag}(\vec{k})]_+=
K\ {\delta}^{\alpha\beta}\ {\delta}_{rs}\ {\delta}^3(\vec{ k}'-\vec{k})
\end{equation}
and
\begin{equation}
{\sum}_r\ u_r(\vec{k})\bar{u}_r(\vec{k})\ =\  N\ 
[\ m_b(1+v{\slash})+k{\slash}\ ] \\ 
\end{equation}
where the normalization factors must obey the combined relation:  
\cite{dona}
\footnote{Note that $I{\neq}1/(2\pi)^32k^0$ since $k$ and $p$ are
not related through a Lorentz transformation}
\begin{equation} 
I\ =\ \frac{K\ N}{J^2}\ =\ \frac{1}{(2\pi)^3\ 2p^0}\ =\ 
\frac{1}{(2\pi)^3\ 2(m_bv^0+k^0)}
\end{equation}
\par
On the other hand, it is quite usual in literature to 
identify effective heavy quark subfields with those ${\lq}{\lq}$leading"
components of the Fourier expansion corresponding to momenta
$p\ {\simeq}\ p_b$ (or equivalently with $k$ components
close to zero), so they rather look like single spinors or anti-spinors
at leading order \vspace{0.1in} in $1/m_b$.
\par 
Instead, in this work I shall deal at first with the full
$\vec{k}$ spectrum, whose components, nevertheless, are supposed to have
a small off-shellness as
mentioned before. Therefore, let us introduce the effective
subfields in the following manner \cite{neub} \cite{mannel}
\footnote{Literature is sometimes not completely clear in their
interpretation as quantum fields. I define the subfields in standard
notation of quantum field theory, where the superscripts $+$ and $-$ label
positive frequencies (associated to annihilation operators of quarks) and
 negative frequencies (associated to creation operators of 
antiquarks) respectively}

\begin{equation}
h_v^{(b,\alpha)+}(x)\ =\ e^{im_bv{\cdot}x}\ \frac{1+v{\slash}}{2}\ 
Q_v^{(b,\alpha)+}(x)
\end{equation}
\begin{equation}
H_v^{(b,\alpha)+}(x)\ =\ e^{im_bv{\cdot}x}\ \frac{1-v{\slash}}{2}\ 
Q_v^{(b,\alpha)+}(x)
\end{equation}
where $H_v^{(b,\alpha)+}(x)$ plays the
role of the ${\lq}{\lq}$small" (lower in the hadron reference frame) 
component of the corresponding spinor subfield. Thus one may identify
from the expansion (4)
\begin{equation}
h_v^{(b,\alpha)+}(x)\ =\  \frac{1+v{\slash}}{2}\ \int\ \frac{d^3\vec{k}}{J}
\ \sum_{r}\ b_r^{\alpha}(\vec{k})\ u_r(\vec{k})\ 
e^{-ik{\cdot}x}
\end{equation}
\begin{equation} 
H_v^{(b,\alpha)+}(x)\ =\ \frac{1-v{\slash}}{2}\ \int\
\frac{d^3\vec{k}}{J}\ \sum_{r}\ b_r^{\alpha}(\vec{k})\ 
u_r(\vec{k})\ e^{-ik{\cdot}x} 
\end{equation} 
\vspace{0.1in}
\par 
The relation \cite{falk}
\[ H_v^{(b,\alpha)+}(x)\ =\ 
\frac{1-v{\slash}}{2}\ \frac{k\slash}{2m_b}\ h_v^{(b,\alpha)+}(x) \]
is verified at leading order, following from the above \vspace{0.1in} 
definitions.
\par
Next I shall derive the explicit space-time representation of
the Feynman propagator for the $h_v^{(b,\alpha)+}$ subfield.
From the relations (7-9) one may easily
find \vspace{0.1in} that \footnote{In order to avoid a large
and misleading notation
the extra minus sign as a superscript in the $\bar{h}_v^+$
subfield has been omitted which, however, should be implicitly
understood since
negative frequencies and creation operators are involved. The same omission
will occur afterwards for the $\bar{H}_v^+$ subfield as well}  
\begin{eqnarray}
iS_{h_v}^+(x-y) & = & 
<0{\mid}h_v^{(b,\alpha)+}(x)\bar{h}_v^{(b,\beta)+}(y){\mid}0> \nonumber \\
& = & {\delta}^{\alpha\beta}\  
\frac{1+v{\slash}}{2}\ \biggl(\ \int\ \frac{d^3\vec{k}}{I}\ 
[(1+v{\slash})m_b+k{\slash})]\ e^{-ik{\cdot}(x-y)}\  \biggr)\ 
\frac{1+v{\slash}}{2}
\end{eqnarray}
where the invariant quantity $I=KN/J^2$ is given by Eq. \vspace{0.1in} (9).
\par
With the aid of the gamma commutation relations and the fact that
$(1{\pm}v{\slash})/2$ are projectors, one obtains easily that
\begin{equation}
iS_{h_v}^+(x-y)\ =\ {\delta}^{\alpha\beta}\ \frac{1+v{\slash}}{2}\ 
\biggl(\ \int\ \frac{d^3\vec{k}}{I}\
[2m_b+k{\cdot}v]\ e^{-ik{\cdot}(x-y)}\  \biggr) 
\end{equation}
\par
Using the Dirac delta properties the above result can be expressed
as a four-dimensional integral
\begin{equation}
iS_{h_v}^+(x-y)\ = \ {\delta}^{\alpha\beta}\ \frac{1+v{\slash}}{2}\ 
\biggl(\ \int\ \frac{d^4k}{(2\pi)^3}\
\theta(k{\cdot}v)\ \delta(k^2+2m_bv{\cdot}k-{\Delta}^2)\ 
[2m_b+k{\cdot}v]\ e^{-ik{\cdot}(x-y)}\  \biggr) \nonumber
\end{equation} 
\par
In the infinite mass limit the kinetic energy of the heavy quark
inside the hadron will be neglected but not its three-momentum
$\vec{k}$. (This is consistent with the space-like character
of the four-momentum $k$.) In effect, I shall assume from Eq. (6)
for the positive root $k^0$,
\begin{equation}
\frac{k{\cdot}v}{2m_b}\ =\ \frac{k^0}{2m_b}\ {\simeq}\ 
\frac{\vec{k}^2+{\Delta}^2}{4m_b^2}\ 
{\rightarrow}\ 0
\end{equation}
\par
In the real world the above limit makes sense for
$\vec{k}^2<<m_b^2$. Neglecting those Fourier components above
$m_b$, which acts as an ultraviolet cutoff, probably makes sense for
the bottom quark but not for the charm \vspace{0.1in} quark.
\par
The two roots of Eq. (5) showed in expression (6) correspond
to the two poles in the real $k^0$-axis of the integrand
of Eq. (16). In the $m_b{\rightarrow}\infty$ limit both poles are shifted to 
{\em zero} and to $-\infty$ respectively. The theta function ensures 
that only the former one contributes, thereby implying
\begin{equation}
iS_{h_v}^+(x-y)\ = \ {\delta}^{\alpha\beta}\ \frac{1+v{\slash}}{2}\ 
\biggl(\ \int\ \frac{d^4 k}{(2\pi)^3}\ 
\delta(2m_bv{\cdot}k)\ 
[2m_b+k{\cdot}v]\ e^{-ik{\cdot}(x-y)}\  \biggr) 
\end{equation} 
\par
Finally, making use of the property:  
$\delta(2m_bv{\cdot}k)=\delta(v{\cdot}k)/2m_b$, one gets
for the Feynman ${\lq}{\lq}$propagator" at leading order
 
\begin{equation}
iS_F^+(x-y)\ =\ \theta(v{\cdot}x-v{\cdot}y)\ iS_{h_v}^+(x-y)\ =\ 
\frac{1+v{\slash}}{2}\
\theta(v{\cdot}x-v{\cdot}y)\ 
\delta^{\alpha\beta}\ {\delta}_v^{3}(\vec{x}-\vec{y}) 
\end{equation}
where 
\[ {\delta}_v^{3}(\vec{x}-\vec{y})\ =\ \int\ \frac{d^4k}{(2\pi)^3}\ 
\delta(k{\cdot}v)\ e^{-ik{\cdot}(x-y)} \]
\par
Observe that there is only a single pole in the
real $k^0$-axis at $k^0=0-i\epsilon$, 
leading to
\begin{equation}
S_F^+(x-y)\ =\ \frac{1+v{\slash}}{2}\ \delta^{\alpha\beta}\
 \int\ \frac{d^4k}{(2\pi)^4}\ \frac{1}{k{\cdot}v+i\epsilon}\ 
e^{-ik{\cdot}(x-y)}
\end{equation} 
\par
The Dirac delta in (19) reflects the fact that the heavy quark remains
immovable inside the hadron moving with four-velocity $v$, as
already pointed out in Ref. \cite{hill}. Finally, I remark that the
propagator in momentum space indeed shows the usual form in 
HQET \cite{grins} \footnote{Obviously  the same expression can
be obtained directly from the equation satisfied at leading order by 
the propagator in HQET:
\[ iv{\cdot}{\partial}\ S_F^+(x-y)\ =\ \frac{1+v{\slash}}{2}\ \delta^4(x-y) \]
and then making use of the Fourier transform}
\begin{equation}
S_F^+(k)\ =\ \frac{1+v{\slash}}{2}\ \delta^{\alpha\beta}\ 
\frac{1}{k{\cdot}v+i\epsilon}
\end{equation}
\par
In a similar way one gets for the propagator corresponding to
the $H_v^{(b,\alpha)+}$ subfield for example,
\begin{equation}
iS_{H_v}^+(x-y)\ =\  
<0{\mid}H_v^{(b,\alpha)+}(x)\bar{H}_v^{(b,\beta)+}(y){\mid}0>\ {\equiv}\ 0
\end{equation}

technically coming from the fact that the following matrix element:
\begin{equation}
\frac{1-v{\slash}}{2}\ \frac{[(1+v{\slash})m+k{\slash}]}{2m_b}\ 
\frac{1-v{\slash}}{2}\ =\  -\frac{1-v{\slash}}{2}\ \frac{k{\cdot}v}{2m_b} 
\end{equation}
vanishes in the $m_b{\rightarrow}0$ limit. If we do not neglect
$(\vec{k}^2+{\Delta}^2)/4m_b^2$ the expression (22) is not identically zero, 
showing explicitly that indeed the ${\lq}{\lq}$small" components
of the effective heavy quark field yield $1/m_b$ \vspace{0.1in} effects. 
\par
Vacuum expectation values of the combined
products of the $h_v^{(b,\alpha)+}$ and $H_v^{(b,\beta)+}$ subfields
lead to further contributions
\footnote{The combinations $<0{\mid}h_v^+\overline{H}_v^+{\mid}0>$ and
$<0{\mid}H_v^+\overline{h}_v^+{\mid}0>$ yield respectively
\[ \frac{1{\pm}v{\slash}}{2}\ 
\frac{k{\slash}\ {\mp}\ k{\cdot}v}{k^2+2m_bk{\cdot}v-{\Delta}^2} \]
in momentum space. In the sum of all four contributions the full
propagator is recovered}
to the full propagator similarly suppressed. Therefore, at leading 
order the only ${\lq}{\lq}$propagator" for a heavy quark is that coming
from the $h_v^{(b,\alpha)+}$ subfield as \vspace{0.2in} expected.
\newline 
{\bf Anti-Particle \vspace{0.1in} Sector}
\par
Proceeding in a parallel way as in the particle sector, let us remark
however that now the Fourier expansion of $Q_v^{(b,\alpha)-}(x)$
involves negative frequencies. Therefore, starting from
Eq. (1) I shall introduce the Fourier residual momentum in this case as
\begin{equation}
p\ =\ m_bv\ -\ k
\end{equation}
so $k^0$ will explicitly exhibit its negative sign. In effect, let
us write
\begin{equation}
Q_v^{(b,\alpha)-}(x)\ =\ 
\ e^{im_bv{\cdot}x}\ \int\   
\frac{d^3\vec{k}}{J}\ \sum_r\ \tilde{b}_r^{\alpha\dag}(\vec{k})\ 
v_r(\vec{k})\ e^{-ik{\cdot}x}
\end{equation}
\par
The slightly off-shellness condition (2) now \vspace{0.1in} implies
from Eq. (24)
\begin{equation}
(k^0)^2\ -\ 2m_bk^0\ -\ \vec{k}^2\ =\ {\Delta}^2
\end{equation}
whose roots are
\vspace{0.1in}
\begin{equation}
k^0\ =\ {\pm}\sqrt{m_b^2+\vec{k}^2+{\Delta}^2}\ + m_b
\end{equation}
but only the negative solution will contribute and
a purely advanced Green function in space-time will appear
at the end of the \vspace{0.1in} calculation.
\par
The effective subfields are defined as

\begin{equation}
h_v^{(b,\alpha)-}(x)\ =\ e^{-im_bv{\cdot}x}\ \frac{1-v{\slash}}{2}\ 
Q_v^{(b,\alpha)-}(x)=\  \frac{1-v{\slash}}{2}\ \int\ \frac{d^3\vec{k}}{J}
\ \sum_{r}\ \tilde{b}_r^{\alpha\dag}(\vec{k})\ v_r(\vec{k})\ 
e^{-ik{\cdot}x}       
\end{equation}
\begin{equation}
H_v^{(b,\alpha)-}(x)\ =\ e^{-im_bv{\cdot}x}\ \frac{1+v{\slash}}{2}\ 
Q_v^{(b,\alpha)-}(x)\ =\ \frac{1+v{\slash}}{2}\ \int\
\frac{d^3\vec{k}}{J}\ \sum_{r}\ \tilde{b}_r^{\alpha\dag}(\vec{k})\ 
v_r(\vec{k})\ e^{-ik{\cdot}x} 
\end{equation}
\vspace{0.1in}
where the anti-spinors satisfy
\begin{equation}
{\sum}_r\ v_r(\vec{k})\bar{v}_r(\vec{k})\ =\  
 N\ [\ m_b(v{\slash}-1)-k{\slash}\ ] 
\end{equation}
\par
Following the same steps and limits as in the particle sector, one obtains
for the negative-frequency propagator
\begin{equation}
iS_{h_v}^-(x-y)\ =\  
<0{\mid}\bar{h}_v^{(b,\alpha)-}(y)h_v^{(b,\beta)-}(x){\mid}0>\ =\ 
-\frac{1-v{\slash}}{2}\ \delta^{\alpha\beta}\ 
{\delta}_v^{3}(\vec{x}-\vec{y}) 
\end{equation}
whereas the propagators involving the $H_v^{(b,\alpha)-}$ subfield do not
contribute at leading order in analogy to Eq. (22). Therefore
the Feynman propagator for a massive anti-quark is
                                                   
\begin{equation}
iS_F^-(x-y)\ =\ -\ \theta(v{\cdot}y-v{\cdot}x)\ iS_{h_v}^-(y-x)\ =\ 
\frac{1-v{\slash}}{2}\ 
\theta(v{\cdot}y-v{\cdot}x)\  
\delta^{\alpha\beta}\ {\delta}_v^{3}(\vec{x}-\vec{y}) 
\end{equation}
where the single pole lies now at $k{\cdot}v=0+i\epsilon$, 
leading to
\begin{equation}
S_F^-(x-y)\ =\ \frac{v{\slash}-1}{2}\ \delta^{\alpha\beta}\ 
\int\ \frac{d^4k}{(2\pi)^4}\ \frac{1}{k{\cdot}v-i\epsilon}\ 
e^{-ik{\cdot}(x-y)}
\end{equation}
\par
Let us make a final remark on the single poles of the propagators. 
Since the heavy quark mass was assumed infinite, the particle and
anti-particle sectors of the effective theory actually decouple
(for example no heavy quark pair production is allowed). 
This manifests in the fact that there is no single expression
representing altogether the propagation forward in time of the positive
energy solutions and backward
in time of the negative energy solutions as otherwise happens in the Dirac 
\vspace{0.1in} theory.
\par
In summary, in this work I have examined 
those effective subfields commonly introduced in HQET to deal
with almost on-shell heavy quarks from the standard
viewpoint of a quantum field theory (although especially
simple in this case). For example, the $\overline{h}_v^+(x)$ field
as defined in Eq. (12) creates a heavy quark at point $x$ in space-time 
as a superposition of
single-particle states with momenta ranging over the entire
$\vec{k}$ domain (actually for $\vec{k}^2<m_b^2$) and not yet limited 
to small values of the components, of the order
of \vspace{0.1in} ${\Lambda}_{QCD}$.\par
However, in constructing the effective theory for heavy quarks one
should add a further restriction on the Fourier expansion of the
fields, limiting the range of the  $\vec{k}$ components to values of the 
order of ${\Lambda}_{QCD}$. Note that this condition amounts to 
a new constraint not completely equivalent to the small virtuality 
already imposed by means of Eq. (2). Indeed, the smallness of 
$\vec{k}$ implies the almost on-shellness condition but the converse is
\vspace{0.1in} not necessarily true.
\par
In fact, eliminating those components with large residual momentum 
in the Fourier expansion of the fields is equivalent to 
integrating away the high-velocity states in the functional integral
formulation of Ref. \cite{italia}. It is also interesting to mention 
that, once restricted the $\vec{k}$
range to values sharply peaked at the origin, the propagator
would spread out over coordinate space with typical
width ${\simeq}\ 1/{\Lambda}_{QCD}$ as a consequence of the
Fourier \vspace{0.2in} transform. \newline
\vspace{0.1in} {\bf Acknowledgments}
\newline
I thank V. Gim\'enez and M. Neubert for reading the manuscript and comments.
\end{document}